\documentclass[prd,twocolumn,tightenlines,showpacs,nofootinbib]{revtex4-1} 

\usepackage{amsfonts}
\usepackage{dcolumn}
\usepackage{bm}
\usepackage{amsmath}
\usepackage{nicefrac}
\usepackage{amsthm}
\usepackage{amssymb}
\usepackage{graphicx}
\usepackage{color}
\usepackage{url}
\usepackage{cancel}
\usepackage{rotating}

\usepackage[colorlinks=true,linkcolor=black,citecolor=blue,urlcolor=blue,bookmarksopen]{hyperref}

\newcommand{\appropto}{\mathrel{\vcenter{
  \offinterlineskip\halign{\hfil$##$\cr
    \propto\cr\noalign{\kern2pt}\sim\cr\noalign{\kern-2pt}}}}}
\renewcommand{\v}[1]{\boldsymbol{#1}}		

\begin{document}

\title{Probing ``long-range'' neutrino-mediated forces with atomic and nuclear spectroscopy}

\date{\today}
\author{Yevgeny~V.~Stadnik} 
\affiliation{Helmholtz Institute Mainz, Johannes Gutenberg University of Mainz, 55128 Mainz, Germany}

\begin{abstract}
The exchange of a pair of low-mass neutrinos between electrons, protons and neutrons produces a ``long-range'' $1/r^5$ potential, which can be sought for in phenomena originating on the atomic and sub-atomic length scales. 
We calculate the effects of neutrino-pair exchange on transition and binding energies in atoms and nuclei. 
In the case of atomic $s$-wave states, there is a large enhancement of the induced energy shifts due to the lack of a centrifugal barrier and the highly singular nature of the neutrino-mediated potential. 
We derive limits on neutrino-mediated forces from measurements of the deuteron binding energy and transition energies in positronium, muonium, hydrogen and deuterium, as well as isotope-shift measurements in calcium ions. 
Our limits improve on existing constraints on neutrino-mediated forces from experiments that search for new macroscopic forces by 18 orders of magnitude. 
Future spectroscopy experiments have the potential to probe long-range forces mediated by the exchange of pairs of standard-model neutrinos and other weakly-charged particles. 
\end{abstract}


\maketitle

\textbf{Introduction.} --- 
The exchange of a pair of neutrinos between two particles is predicted to mediate a long-range force between the particles \cite{Brown1996,Feinberg1968}. 
In Ref.~\cite{Feinberg1968} (see also \cite{Feinberg1989,Sikivie1994,Grifols1996,Grifols1999,Shuryak2007,Lusignoli2011}), the long-range part of the potential due to the exchange of a pair of massless neutrinos between two particles was calculated and found to scale as $\propto 1/r^5$.  
The $1/r^5$ neutrino-mediated potential induces a feeble $1/r^6$ force which is far too small to detect with current experiments that search for new macroscopic forces \cite{Adelberger2007A,Adelberger2007B,Decca2016,Romalis2009,Adelberger2015}. 
The implications of many-body neutrino-mediated forces in stars were considered in Ref.~\cite{Fischbach1996}, but it was subsequently pointed out that the effects of such forces are suppressed in all types of stars \cite{Smirnov1996}.

In the present work, we consider the novel approach of searching for effects associated with the neutrino-mediated $1/r^5$ potential on atomic and sub-atomic length scales. 
Measurements of transition and binding energies in atoms and nuclei provide a powerful way of probing neutrino-mediated forces, since energy differences (or equivalently frequencies) are among the most accurately measurable physical quantities. 
In particular, current state-of-the-art atomic and ionic clocks operating on optical transitions have demonstrated a fractional accuracy approaching the level of $\sim 10^{-18}$ \cite{Ye2014,Katori2015,Ye2015,Peik2016}.

Atomic $s$-wave states (states with orbital angular momentum $l=0$) offer an ideal platform to search for the neutrino-mediated $1/r^5$ potential due to the lack of a centrifugal barrier and the highly singular nature of the $1/r^5$ potential. 
As the simplest example, the radial part of the non-relativistic $1s$ hydrogen wavefunction scales as $R_{1s} (r) \propto r^0/a_\textrm{B}^{3/2}$ at small distances (where $a_\textrm{B} \approx 5.29 \times 10^{-11}~\textrm{m}$ is the atomic Bohr radius), meaning that the integral $\int_{r_c}^\infty r^2 [R_{1s}(r)]^2 /r^5 dr$ diverges like $\propto 1/r_c^2$ for a point-like nucleus ($r_c = 0$). 
In the physical hydrogen atom with a finite-size nucleus ($r_c \sim 10^{-15}~\textrm{m}$), this integral is finite and scales parametrically like $\sim 1/(a_\textrm{B}^3 r_c^2)$, which is enhanced compared to the characteristic $\sim 1/a_\textrm{B}^5$ scaling in atomic states of higher orbital angular momentum by the factor $(a_\textrm{B}/r_c)^2 \sim 10^9$.


\textbf{Potential induced by the exchange of a pair of low-mass neutrinos.} --- 
The potential mediated by the exchange of a pair of neutrinos of non-zero mass $m_\nu$ is long range in an atom, if the size of the atom is much smaller than the Yukawa range parameter associated with the pair of neutrinos, $\lambdabar = 1/(2 m_\nu) \gg R_\textrm{atom}$. 
Beta-decay experiments directly constrain the electron-antineutrino mass to be $m_{\bar{\nu}_e} \lesssim 2~\textrm{eV}$ \cite{Mainz2005,Troitsk2011}, while cosmological observations give model-dependent constraints on the sum of the three different neutrino masses at a comparable level \cite{PDG2017}. 
This implies a Yukawa range parameter of $\lambdabar \gtrsim 10^{-7}~\textrm{m} \gg R_{\textrm{atom}}$, and so we can treat the neutrino-mediated potential as being long range. 

The long-range part of the potential due to the exchange of a pair of low-mass neutrinos between two fermions reads as follows \cite{Feinberg1968,Feinberg1989,Sikivie1994}: 
\small
\begin{align}
\label{Neutrino_potential_master}
V_{\nu}(\v{r}) = &\frac{G_\textrm{F}^2}{4 \pi^3 r^5} \left\{a_1 a_2 
- b_1 b_2 \left[ \frac{3}{2} \v{\sigma}_1 \cdot \v{\sigma}_2 - \frac{5}{2} \left(\v{\sigma}_1 \cdot \v{\hat{r}}\right) \left(\v{\sigma}_2 \cdot \v{\hat{r}}\right) \right] \right\} \notag \\
&= \frac{G_\textrm{F}^2}{4 \pi^3 r^5} \left\{a_1 a_2 
- \frac{2}{3} b_1 b_2  \v{\sigma}_1 \cdot \v{\sigma}_2 \right. \notag \\
&- \left. \frac{5}{6} b_1 b_2 \left[ \v{\sigma}_1 \cdot \v{\sigma}_2 - 3 \left(\v{\sigma}_1 \cdot \v{\hat{r}}\right) \left(\v{\sigma}_2 \cdot \v{\hat{r}}\right) \right] \right\} \, ,
\end{align}
\normalsize
where $G_\textrm{F} \approx 1.166 \times 10^{-5}~\textrm{GeV}^{-2}$ is the Fermi constant of the weak interaction, $r$ is the distance between the two fermions, $\v{\sigma}_1$ and $\v{\sigma}_2$ are the Pauli spin matrix vectors of fermions 1 and 2, and $\v{\hat{r}}$ is the unit vector directed between the two fermions. 
In the present work, we focus mainly on systems in $l=0$ states, which are described by spherically symmetric wavefunctions. 
For such states, the expectation value of the rank-2 tensor part in (\ref{Neutrino_potential_master}) vanishes:~$\left< [ \v{\sigma}_1 \cdot \v{\sigma}_2 - 3 \left(\v{\sigma}_1 \cdot \v{\hat{r}}\right) \left(\v{\sigma}_2 \cdot \v{\hat{r}}\right) ] / r^5 \right>_{l=0} = 0$. 

The species-dependent parameters $a_i$ and $b_i$ in Eq.~(\ref{Neutrino_potential_master}) are determined by several processes involving weak neutral and charged currents (see Fig.~1 of Ref.~\cite{Sikivie1994}). 
For a single neutrino species, charged leptons receive contributions from both the weak neutral and charged currents:~$a^{(1)}_l = 1 + g_l^V = 1/2 + 2 \sin^2(\theta_\textrm{W})$ and $b^{(1)}_l = 1 + g_l^A = 1/2$, with $\sin^2(\theta_\textrm{W}) \approx 0.24$ \cite{Footnote1}, while nucleons receive a contribution solely from the weak neutral currents:~$a^{(1)}_n = -1/2$, $a^{(1)}_p = 1/2 - 2 \sin^2(\theta_\textrm{W}) $, $b^{(1)}_n = -g_A/2$, and $b^{(1)}_p = g_A/2$, with $g_A \approx 1.27$. 
The nucleons and charged leptons also receive contributions from the other two neutrino species, due purely to the weak-neutral-current processes, with each neutrino species contributing the amount $a^{(2)}_l = g_l^V$, $b^{(2)}_l = g_l^A$, $a^{(2)}_N = a^{(1)}_N$ and $b^{(2)}_N = b^{(1)}_N$. 

Furthermore, there are additional contributions from other weakly-charged species (species that participate in weak processes) of mass $m$ via the purely weak-neutral-current process, when the dominant effects of the potential (\ref{Neutrino_potential_master}) arise at length scales $L \ll 1/(2m)$. 
The effects of these weakly-charged species are analogous to the effects of neutrino species in the weak-neutral-current channel, except for an overall numerical constant, which is given by $2[ (g_l^V)^2 + (g_l^A)^2 ] \approx 0.501$ for a charged lepton species, $2[ (g_u^V)^2 + (g_u^A)^2 ] \approx 0.565$ for an up-type quark species, and $2[ (g_d^V)^2 + (g_d^A)^2 ] \approx 0.731$ for a down-type quark species. 
These numerical constants are normalised to the value for a neutrino species:~$2[ (g_\nu^V)^2 + (g_\nu^A)^2 ] = 1$. 

Altogether, the combinations of species-dependent parameters in Eq.~(\ref{Neutrino_potential_master}) therefore have the following effective values: 
\begin{align}
\label{Eq1}
a_1 a_2 \to a_1^{(1)} a_2^{(1)} + (N_\textrm{eff} - 1) a_1^{(2)} a_2^{(2)} \, , \\ 
\label{Eq2}
b_1 b_2 \to b_1^{(1)} b_2^{(1)} + (N_\textrm{eff} - 1) b_1^{(2)} b_2^{(2)} \, ,  
\end{align}
where $N_\textrm{eff}$ is the effective number of neutrino species. 
In systems where the dominant effects of (\ref{Neutrino_potential_master}) arise on the atomic length scale, the main contributions are from the species $\nu_e$, $\nu_\mu$ and $\nu_\tau$, giving $N_\textrm{eff} \approx 3$. 
In systems where the dominant effects of (\ref{Neutrino_potential_master}) arise on the nuclear length scale, the main contributions are from the species $\nu_e$, $\nu_\mu$, $\nu_\tau$ and $e$, giving $N_\textrm{eff} \approx 3.50$ \cite{Footnote2}. 
Finally, in systems where the dominant effects of (\ref{Neutrino_potential_master}) arise on a length scale of the order of the Compton wavelength of the $Z$ or $W$ boson, the main contributions are from the species $\nu_e$, $\nu_\mu$, $\nu_\tau$, $e$, $\mu$, $\tau$, $u$, $c$, $d$, $s$ and $b$, giving $N_\textrm{eff} \approx 14.47$, taking into account that each quark has three possible colours. 
The overall sign of the potential (\ref{Neutrino_potential_master}) is reversed when one of the two fermions is replaced by its antiparticle.

\textbf{Deuteron binding energy.} --- 
Deuteron --- the bound state of a proton and a neutron in the $^3 S_1$ state (with a small admixture of the $^3 D_1$ state, which can be neglected in the first approximation) --- can be simply modelled by a spherical potential well with an infinitely repulsive inner hard core, in which the potential between the two nucleons takes the following form: 
\begin{eqnarray}
\label{spherical_hardcore_potential}
V_\textrm{nucl}(r) = \left\{ \begin{array}{ll}
+\infty & \textrm{~for $r < r_1$,}\\
-|V_0| & \textrm{~for $r_1 < r < r_2$,}\\
0 & \textrm{~for $r > r_2$,}
\end{array} \right.
\end{eqnarray}
where $|V_0|$ is the depth of the spherical potential well. 
For our estimates below, we assume the values $r_1 = 0.5~\textrm{fm}$ and $r_2 = 2.5~\textrm{fm}$. 

The radial wavefunction solutions of the potential (\ref{spherical_hardcore_potential}) for an $s$-wave state are given by: 
\begin{eqnarray}
\label{spherical_hardcore_solutions}
R_{s}(r) = \left\{ \begin{array}{ll}
0 & \textrm{~for $r \le r_1$,}\\
C_1 j_0 \left( kr \right) + C_2 n_0 \left( kr \right) & \textrm{~for $r_1 \le r \le r_2$,}\\
C_3 h_0^{(1)} (i \kappa r) & \textrm{~for $r \ge r_2$,}
\end{array} \right.
\end{eqnarray}
where $j$ is the spherical Bessel function of the first kind, 
$n$ is the spherical Bessel function of the second kind, 
$h^{(1)}$ is the spherical Hankel function of the first kind, 
$k = \sqrt{2 \mu (|V_0| - E_B)}$ 
and $\kappa = \sqrt{2 \mu E_B}$, 
with $\mu = m_n m_p / (m_n + m_p) \approx 0.47~\textrm{GeV}$ being the deuteron reduced mass and $E_B \approx 2.2~\textrm{MeV}$ the deuteron binding energy. 
Requiring the continuity of $R_s$ at $r=r_1$ and the continuity of both $R_s$ and $dR_{s}/dr$ at $r=r_2$, we determine that $|V_0| \approx 36~\textrm{MeV}$. 
The normalisation condition $\int_0^\infty r^2 |R_{s}(r)|^2 dr = 1$ then fixes the normalisation constants in Eq.~(\ref{spherical_hardcore_solutions}) to be $C_1 \approx 0.46~\textrm{fm}^{-3/2}$, $C_2 \approx 0.22~\textrm{fm}^{-3/2}$, and $C_3 \approx - 0.22~\textrm{fm}^{-3/2}$. 

Using the wavefunction in Eq.~(\ref{spherical_hardcore_solutions}), we calculate the expectation value of the $1/r^5$ operator for the deuteron bound state to be: 
\begin{equation}
\label{expectation_value_1/r^5_deuteron}
\left< \left. ^3 S_1 \right. \left| \frac{1}{r^5} \right|  \left.^3 S_1 \right. \right> \approx 0.060~\textrm{fm}^{-5} \, . 
\end{equation}
Using the result (\ref{expectation_value_1/r^5_deuteron}), we determine the change in the deuteron binding energy due to the neutrino-mediated potential (\ref{Neutrino_potential_master}) to be: 
\begin{equation}
\label{expectation_value_deuteron_FULL}
\delta E_B (^3 S_1) \approx - \frac{ G_\textrm{F}^2}{4 \pi^3} \left(a_n a_p - \frac{2}{3} b_n b_p \right) \times 0.060~\textrm{fm}^{-5} \, . 
\end{equation}

Comparing the measured \cite{Deuteron_Exp1999} and predicted \cite{Deuteron_Theor2003,Deuteron_Theor2015} values of the deuteron binding energy:
\begin{align}
\label{D_BE_exp}
E_B^\textrm{exp} = 2.2245663(4)~\textrm{MeV} \, , \\
\label{D_BE_theor}
E_B^\textrm{theor} = 2.22457(1)~\textrm{MeV} \, ,
\end{align}
and using expressions (\ref{expectation_value_deuteron_FULL}), (\ref{Eq1}) and (\ref{Eq2}), we place the following constraint on the neutrino-mediated potential in Eq.~(\ref{Neutrino_potential_master}): 
\begin{equation}
\label{deuteron_BE_result}
G_\textrm{eff}^2 \lesssim 7.9 \times 10^8 ~ G_\textrm{F}^2 \, . 
\end{equation}

\textbf{Spectroscopy of simple atoms.} --- 
Simple two-body atoms with relatively light nuclei ($Z \sim 1$) can be treated in the non-relativistic framework. 
Using the non-relativistic form of the wavefunctions for a hydrogen-like system \cite{LL3}, we calculate the expectation value of the $1/r^5$ operator for $l=0$ atomic states to be: 
\begin{equation}
\label{expectation_value_1/r^5_simple-atom}
\left< \left. ns \right. \left| \frac{1}{r^5} \right|  \left. ns \right. \right> \approx \frac{2 Z^3}{n^3 r_c^2 \tilde{a}_\textrm{B}^3} \, , 
\end{equation}
where $n$ is the principal quantum number, $Z$ is the electric charge of the nucleus (in units of the proton electric charge $e$), and $\tilde{a}_\textrm{B}$ is the reduced atomic Bohr radius. 
The cutoff parameter $r_c$ in (\ref{expectation_value_1/r^5_simple-atom}) depends on the specific system. 
In atoms with a hadronic nucleus, $r_c$ is given by the nuclear radius $R_\textrm{nucl}$, while in exotic atoms with a non-hadronic point-like ``nucleus'', $r_c$ is determined by the reduced Compton wavelength of the $Z$ boson, $\lambdabar_Z \approx 2.16 \times 10^{-3}~\textrm{fm}$, which is the length scale below which the Fermi four-fermion approximation is no longer valid and the long-range potential in Eq.~(\ref{Neutrino_potential_master}) changes to a much less singular $1/r$ form.

\emph{Positronium and muonium spectroscopy.} --- 
The absence of hadronic nuclei in positronium (a bound state of an electron and a positron) and muonium (a bound state of an electron and an anti-muon) make these very clean systems to study. 
Using the result (\ref{expectation_value_1/r^5_simple-atom}) with $Z=1$, we determine the energy shifts in the positronium and muonium $n~^3 S_1$ and $n~^1 S_0$ states due to the neutrino-mediated potential (\ref{Neutrino_potential_master}) to be: 
\begin{align}
\label{expectation_value_positronium_FULL_triplet}
\delta E(n~^3 S_1) \approx - \frac{ G_\textrm{F}^2}{4 \pi^3} \frac{2}{ n^3 \lambdabar_Z^2 \tilde{a}_\textrm{B}^3 } \left(a_l^2 - \frac{2}{3} b_l^2 \right) \, , \\ 
\label{expectation_value_positronium_FULL_singlet}
\delta E(n~^1 S_0) \approx - \frac{ G_\textrm{F}^2}{4 \pi^3} \frac{2}{ n^3 \lambdabar_Z^2 \tilde{a}_\textrm{B}^3 } \left(a_l^2 + 2 b_l^2 \right) \, , 
\end{align}
with $\tilde{a}_\textrm{B} = 2 a_\textrm{B}$ in positronium and $\tilde{a}_\textrm{B} \approx a_\textrm{B}$ in muonium.

Comparing the measured \cite{Ps_Exp1993} and predicted \cite{Ps_Theory1999} values of the positronium $1~^3 S_1$ $-$ $2~^3 S_1$ transition frequency:
\begin{align}
\label{Ps_exp}
\nu^\textrm{exp}_{1S - 2S} = 1 233 607 216.4(3.2)~\textrm{MHz} \, , \\
\label{Ps_theor}
\nu^\textrm{theor}_{1S - 2S} = 1 233 607 222.18(58)~\textrm{MHz} \, ,
\end{align}
and using expressions (\ref{expectation_value_positronium_FULL_triplet}), (\ref{Eq1}) and (\ref{Eq2}), we place the following constraint on the neutrino-mediated potential in Eq.~(\ref{Neutrino_potential_master}): 
\begin{equation}
\label{Ps_result}
G_\textrm{eff}^2 \lesssim 2.6 \times 10^{8} ~ G_\textrm{F}^2 \, . 
\end{equation}

Additionally, comparing the measured \cite{Ps_HFS_Exp1984} and predicted \cite{Ps_Theory1999} values of the positronium $1~^1 S_0$ $-$ $1~^3 S_1$ ground-state hyperfine splitting interval:
\begin{align}
\label{Ps_exp_hfs}
\nu^\textrm{exp}_\textrm{hfs} = 203389.10(74)~\textrm{MHz} \, , \\
\label{Ps_theor_hfs}
\nu^\textrm{theor}_\textrm{hfs} = 203392.01(46)~\textrm{MHz} \, ,
\end{align}
and using expressions (\ref{expectation_value_positronium_FULL_triplet}), (\ref{expectation_value_positronium_FULL_singlet}) and (\ref{Eq2}), we place the following constraint on the neutrino-mediated potential in Eq.~(\ref{Neutrino_potential_master}): 
\begin{equation}
\label{Ps_hfs_result}
G_\textrm{eff}^2 \lesssim 1.5 \times 10^{7} ~ G_\textrm{F}^2 \, . 
\end{equation}


Finally, comparing the measured \cite{Muonium_HFS_Exp1999} and predicted \cite{Muonium_HFS_Theory1999,CODATA2014} values of the muonium ground-state hyperfine splitting interval:
\begin{align}
\label{M_hfs_exp}
\nu^\textrm{exp}_\textrm{hfs} = 4463302776(51)~\textrm{Hz} \, , \\
\label{M_hfs_theor}
\nu^\textrm{theor}_\textrm{hfs} = 4463302868(271)~\textrm{Hz} \, ,
\end{align}
and using expressions (\ref{expectation_value_positronium_FULL_triplet}), (\ref{expectation_value_positronium_FULL_singlet}) and (\ref{Eq2}), we place the following constraint on the neutrino-mediated potential in Eq.~(\ref{Neutrino_potential_master}): 
\begin{equation}
\label{M_hfs_result}
G_\textrm{eff}^2 \lesssim 1.9 \times 10^{2} ~ G_\textrm{F}^2 \, . 
\end{equation}
The energy shift in the muonium ground-state hyperfine interval due to the long-range $1/r^5$ interaction mediated by pairs of standard-model neutrinos and other weakly-charged particles is at the level $\approx 2~\textrm{Hz}$.

\emph{Hydrogen and deuterium isotope-shift spectroscopy.} --- 
Using the result (\ref{expectation_value_1/r^5_simple-atom}) with $Z=1$ and $\tilde{a}_\textrm{B} \approx a_\textrm{B}$, we determine the energy shifts in the hydrogen and deuterium $l=0$ states, averaged over the respective hyperfine intervals, due to the neutrino-mediated potential (\ref{Neutrino_potential_master}) to be: 
\begin{equation}
\label{expectation_value_hydrogen-deuterium_FULL_averaged}
\delta E(ns) \approx \frac{ G_\textrm{F}^2}{4 \pi^3} \frac{a_l Q_W}{ n^3 R_\textrm{nucl}^2 a_\textrm{B}^3 } \, , 
\end{equation}
where $Q_W \equiv 2(N a_n + Z a_p)$ is the weak nuclear charge, with $N$ being the neutron number and $Z$ the proton number. 

The measured and predicted differences of the $1s$ $-$ $2s$ transition frequency, averaged over the hyperfine interval, in deuterium and hydrogen are \cite{H+D_Exp+Theor2010}:
\begin{align}
\label{H-D_exp}
\nu^\textrm{D,exp}_{1S - 2S} - \nu^\textrm{H,exp}_{1S - 2S} = 670994334606(15)~\textrm{Hz} \, , \\
\label{H-D_theor}
\nu^\textrm{D,theor}_{1S - 2S} - \nu^\textrm{H,theor}_{1S - 2S} = 670994348.9(4.9)~\textrm{kHz} \, ,
\end{align}
where for the predicted value, we have determined the dominant finite-nuclear-size effect using Eq.~(2) of Ref.~\cite{H+D_Exp+Theor2010}, together with the experimentally determined charge radii of the proton and deuteron from spectroscopy measurements in muonic hydrogen and muonic deuterium:~$r_p = 0.84184(67)~\textrm{fm}$ \cite{Proton_size2010} and $r_d = 2.12562(78)~\textrm{fm}$ \cite{Deuteron_size2016}. 
Comparing the measured and predicted values in (\ref{H-D_exp}) and (\ref{H-D_theor}), and using expressions (\ref{expectation_value_hydrogen-deuterium_FULL_averaged}) and (\ref{Eq1}), we place the following constraint on the neutrino-mediated potential in Eq.~(\ref{Neutrino_potential_master}): 
\begin{equation}
\label{H-D_result}
G_\textrm{eff}^2 \lesssim 1.6 \times 10^{11} ~ G_\textrm{F}^2 \, . 
\end{equation}

\textbf{Spectroscopy of heavy atoms.} --- 
In heavy atoms ($Z \gg 1$), the spin-dependent terms of the potential (\ref{Neutrino_potential_master}) are largely ineffective, compared with the spin-independent term. 
The reason for this is that the spin-independent part of the potential (\ref{Neutrino_potential_master}) acts coherently in atoms and scales roughly with the number of neutrons $N \gg 1$, whereas the spin-dependent part acts incoherently in atoms, since ground-state nuclei have at most two unpaired nucleon spins (due to the nuclear pairing interaction). 

Using the relativistic atomic wavefunctions for a valence electron at small distances \cite{Khriplovich1991Book} (where the Coulomb field of the nucleus is unscreened), we calculate the expectation value of the operator (\ref{Neutrino_potential_master}), due to neutrino-pair exchange between atomic electrons and nucleons, for an atomic single-particle state with total angular momentum $j=1/2$ to be: 
\begin{align}
\label{expectation_value_1/r^5_heavy-atom}
\delta E_\kappa 
\approx &\frac{ G_\textrm{F}^2}{4 \pi^3}
\frac{[(\kappa - \gamma)^2 + (Z \alpha)^2] Z (Z_i + 1)^2}{(2-\gamma) \nu^3 R_\textrm{nucl}^2 a_\textrm{B}^3} \notag \\
&\times \frac{a_l Q_W}{[\Gamma(2 \gamma +1)]^2 } \left( \frac{a_\textrm{B}}{2 Z R_\textrm{nucl}} \right)^{2-2\gamma} 
\, , 
\end{align}
where $\kappa = (-1)^{1-l}$, $\gamma = \sqrt{1-(Z\alpha)^2}$, $\alpha \approx 1/137$ is the electromagnetic fine-structure constant, $Z_i$ is the net charge of the atomic species (for a neutral atom $Z_i = 0$), and $\nu$ is the effective principal quantum number, defined via the ionisation energy of the valence electron:~$I = m_e \alpha^2 (Z_i+1)^2 / (2 \nu^2)$. 
In heavy nuclei, the nuclear radius is generally well described by the relation $R_\textrm{nucl} = A^{1/3} r_0$, where $A = Z+N$ is the nucleon number and $r_0 \approx 1.2~\textrm{fm}$. 

To estimate the contribution of neutrino-pair exchange between atomic electrons to the energy shift in heavy atoms, we note that in this case the valence atomic electrons now interact predominantly with a `core' of two $1s$ electrons (which are situated mainly at the distances $r \sim r_{1s} = a_\textrm{B}/Z$), instead of mainly with the $N$ neutrons of the nucleus. 
Thus the electron-electron contribution to the energy shift in heavy atoms is parametrically suppressed compared to the electron-nucleon contribution by the factor $(R_\textrm{nucl}/r_{1s})^2/N \ll 1$.

\emph{Calcium-ion isotope-shift spectroscopy and non-linearities of the King plot.} --- 
Many-electron atomic systems function as the most precise systems in metrology, with optical atomic and ionic clocks already demonstrating a fractional precision at the level of $\sim 10^{-18}$ \cite{Ye2014,Katori2015,Ye2015,Peik2016}. 
At the same time, the complexity of many-electron atoms means that theoretically predicted values for transition frequencies in these systems generally have a precision that is many orders of magnitude worse than the corresponding experimental precision. 
To circumvent this issue, one can utilise isotope-shift measurements in atoms and look for non-linearities in the King plot \cite{King1963}, a technique which was recently considered in Refs.~\cite{Delaunay2016,Frugiele2016,Berengut2017,Viatkina2017} in the different context of probing Yukawa interactions of hypothetical Higgs-like particles. 

As a specific example, we consider isotope-shift spectroscopy measurements in Ca$^+$ ($Z=20$, $Z_i = 1$). 
Ca$^+$ is an excellent system for isotope-shift spectroscopy, since it has five stable or long-lived isotopes with spinless nuclei ($A = 40, 42, 44, 46, 48$), as well as several readily accessible optical transitions \cite{Schmidt2015,Schmidt2017}. 
We shall consider the pair of transitions, $^2 S_{1/2}$ $-$ $^2 P_{1/2}$ and $^2 D_{3/2}$ $-$ $^2 P_{1/2}$, to which we refer as transitions 1 and 2, respectively. 
In this case, the dominant effect of the neutrino-mediated potential (\ref{Neutrino_potential_master}) is on the $S$-level in transition 1, see Eq.~(\ref{expectation_value_1/r^5_heavy-atom}).  We can thus write the differences in the transition frequency between two isotopes $A$ and $A'$, $\nu_i^{A A'} = \nu_i^A - \nu_i^{A'}$, for the two transitions in the following form: 
\begin{align}
\label{Ca+_1_Shift}
\nu_1^{A A'} \approx K_1 \mu_{A A'} + F_1 \delta \left< r^2 \right>_{AA'} - \delta E_{\kappa = -1}^{AA'} \, , \\
\label{Ca+_2_Shift}
\nu_2^{A A'} \approx K_2 \mu_{A A'} + F_2 \delta \left< r^2 \right>_{AA'}  \, ,
\end{align}
where $K_{i}$ and $F_{i}$ are the usual mass-shift and field-shift parameters, $\mu_{A A'} = 1/m_A - 1/m_{A'}$, with $m_A$ and $m_{A'}$ being the respective masses of isotopes $A$ and $A'$, and $\delta E_{\kappa = -1}^{AA'} = \delta E_{\kappa = -1}^{A} - \delta E_{\kappa = -1}^{A'} $ is the difference of the neutrino-induced $S$-level energy shift between the two isotopes $A$ and $A'$. 
Dividing Eqs.~(\ref{Ca+_1_Shift}) and (\ref{Ca+_2_Shift}) by $\mu_{A A'}$, and simultaneously solving the resulting equations, we can eliminate the difference in the square of the charge radii between the two isotopes, $\delta \left< r^2 \right>_{AA'}$, to give the following equation in terms of the modified frequencies $M \nu_i^{A A'} = \nu_i^{A A'} / \mu_{A A'}$: 
\begin{equation}
\label{Ca+_Modified_Shift}
M \nu_1^{A A'} = K_{12} + F_{12} M \nu_2^{A A'} - \frac{\delta E_{\kappa = -1}^{AA'}}{\mu_{A A'}} \, ,
\end{equation}
where $K_{12} = K_1 - F_{12} K_2$ and $F_{12} = F_1/F_2$.

\begin{figure}[t!]
\begin{center}
\includegraphics[width=8.5cm]{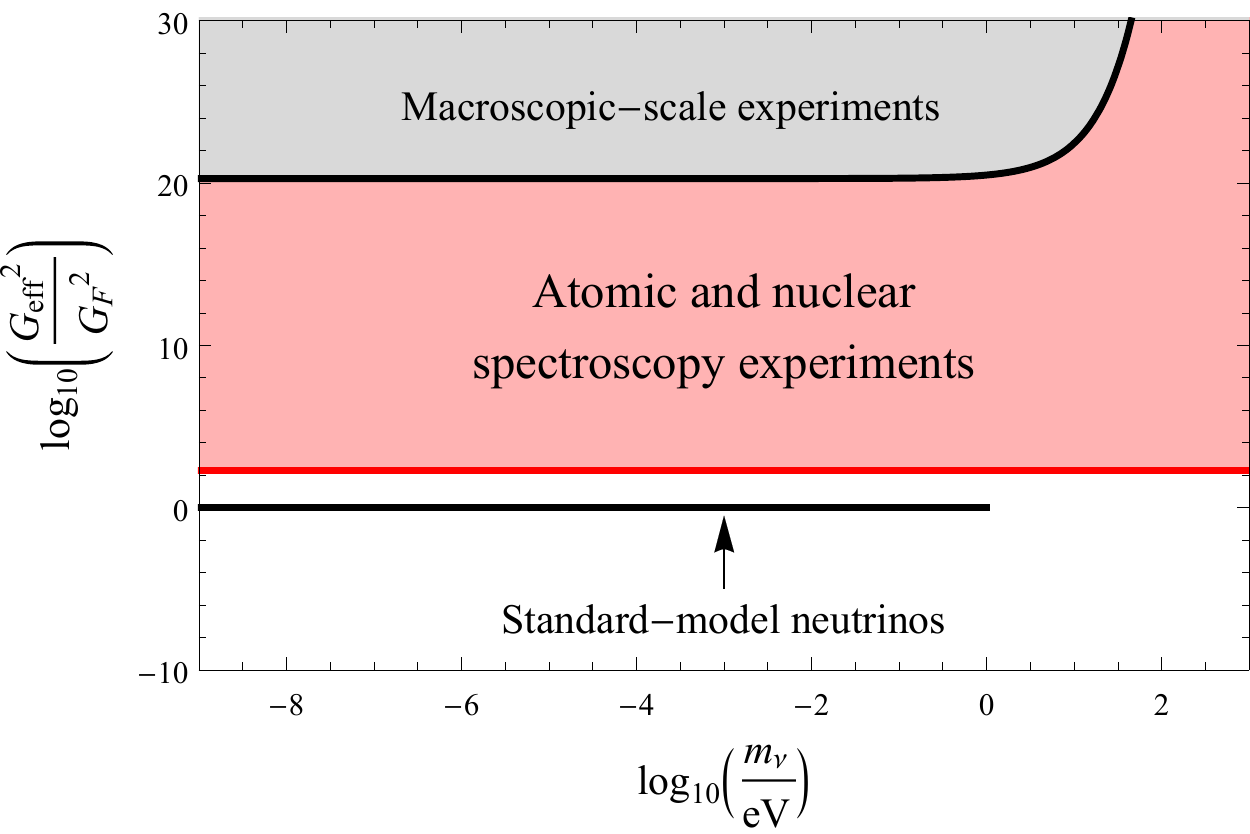} 
\caption{(Color online) 
Limits on the neutrino-mediated potential in Eq.~(\ref{Neutrino_potential_master}), as a function of the neutrino mass $m_\nu$. 
The red region represents constraints derived in the present work from atomic and nuclear spectroscopy. 
The grey region represents constraints from experiments that search for new macroscopic forces \cite{Adelberger2007A,Adelberger2007B,Decca2016,Romalis2009,Adelberger2015}. 
The black line with $G_\textrm{eff} = G_\textrm{F}$ corresponds to the strength of forces mediated by neutrinos and other weakly-charged particles in the standard model. 
} 
\label{fig:Results_summary}
\end{center}
\end{figure}

We see that the last term on the right-hand side of Eq.~(\ref{Ca+_Modified_Shift}) scales as $\propto A A'$ and thus gives rise to a non-linearity in the plot of $M \nu_1^{A A'}$ versus $M \nu_2^{A A'}$ (the so-called King plot \cite{King1963}). 
Such non-linearities have been constrained experimentally at the level $\lesssim 25~\textrm{MHz} \cdot \textrm{GeV}$ over the interval $40 \le A \le 48$ \cite{Schmidt2015}. 
We can use this experimental result, together with expressions (\ref{expectation_value_1/r^5_heavy-atom}) and (\ref{Eq1}), to constrain the neutrino-mediated potential in Eq.~(\ref{Neutrino_potential_master}). 
For the input parameters of (\ref{expectation_value_1/r^5_heavy-atom}), we use the measured value of the ionisation energy of the $^2 S_{1/2}$ state in Ca$^+$:~$I \approx 11.9~\textrm{eV}$ \cite{NIST_database}, and, since the measured differences in the mean-square nuclear charge radii are relatively small across all of the relevant Ca$^+$ isotopes \cite{Schmidt2015}, for simplicity we can assume the nuclear radius $R_\textrm{nucl} = A^{1/3} r_0$ with $A = 44$ for all of the relevant isotopes. 
This yields the following constraint: 
\begin{equation}
\label{Ca+_result}
G_\textrm{eff}^2 \lesssim 4.0 \times 10^{11} ~ G_\textrm{F}^2 \, . 
\end{equation}

\textbf{Conclusions.} --- 
We have calculated the effects of the neutrino-mediated potential in Eq.~(\ref{Neutrino_potential_master}) on transition and binding energies in atoms and nuclei. 
Using existing spectroscopy data, we have derived constraints on neutrino-mediated forces (see Fig.~\ref{fig:Results_summary}). 
Our derived limits improve on existing constraints on neutrino-mediated forces from experiments that search for new spin-independent \cite{Adelberger2007A,Adelberger2007B,Decca2016} and spin-dependent \cite{Romalis2009,Adelberger2015} macroscopic forces by 18 orders of magnitude. 

With a sufficient improvement in experimental and theoretical precision, future spectroscopy experiments have the potential to probe long-range forces mediated by the exchange of pairs of standard-model neutrinos and other weakly-charged particles. 
The observation of neutrino-mediated forces via atomic spectroscopy requires only a single interval. 
The most promising interval at the moment appears to be the ground-state hyperfine interval in muonium. 
The theoretical precision of this interval is currently limited by an independent experimental determination of the electron-to-muon mass ratio \cite{Muonium_HFS_Theory1999,CODATA2014}. 
The purely theoretical uncertainties in this case are all sub-leading \cite{Muonium_HFS_Theory1999,CODATA2014} and, with the exception of fourth-order QED processes (where some terms still need to be calculated), are either smaller than or comparable to the size of the frequency shift expected from neutrino-mediated forces in the standard model. 
In order to probe neutrino-mediated forces within the standard model, one will require a more precise and complete calculation of fourth-order QED contributions, as well as calculations of fifth-order QED contributions and all other one-loop electroweak contributions that do not involve neutrinos.


\textbf{Acknowledgements.} --- 
I am grateful to Victor Flambaum for helpful discussions. 
This work was supported by the Humboldt Research Fellowship.




\begin{thebibliography}{99}
\bibitem{Footnote1} The effects of the running of the Weinberg angle parameter $\sin^2(\theta_\textrm{W})$ are relatively small for the momentum scales that are relevant in the present work \cite{Marciano2000}, so we assume the constant value $\sin^2(\theta_\textrm{W}) \approx 0.24$.
\bibitem{Footnote2} We expect the contribution from strongly-interacting species to be relatively small in this case, since at low momenta transfer quarks form hadrons, which mediate relatively short-ranged interactions.
\end{thebibliography}
\end{document}